\documentclass[aip,jcp,reprint,showpacs]{revtex4-1}
\usepackage{graphicx,amsfonts,amsmath,amssymb,color}
\usepackage[english]{babel}

\newcommand{\p}[2]{\frac{\partial\, #1}{\partial\, #2}\,}

\newcommand{\V}[1]{\mathbf{#1}}
\newcommand{\intl}[2]{\int\limits_{#1}^{#2}}
\newcommand{\bracket}[1]{\left(#1\right)}

\newcommand{\eq}[1]{$\mathrm{Eq.}$~\eqref{#1}}
\newcommand{\figref}[1]{$\mathrm{Fig.}$~\ref{#1}}

\newcommand{\av}[1]{\left\langle #1 \right\rangle}

\begin{document}

\title{Impact of inertia on biased Brownian transport in confined geometries}

\author{S. Martens}
\email{steffen.martens@physik.hu-berlin.de}
\author{I. M. Sokolov}
\author{L. Schimansky-Geier}
\affiliation{Department of Physics, Humboldt-Universit\"{a}t zu Berlin, Newtonstr. 15, 12489 Berlin, Germany}

\begin{abstract}
\noindent We consider the impact of inertia on biased Brownian motion of point particles in a two-dimensional channel with sinusoidally varying width. 
If the time scales of the problem separate, the adiabatic elimination of the transverse degrees of freedom leads to an effective description 
for the motion along the channel given by the potential of mean force. The possibility of such description is intimately connected with equipartition. 
Numerical simulations show that in the presence of external bias the equipartition may break down leading to non-monotonic dependence of mobility on 
external force and several other interesting effects. 
\end{abstract}

\maketitle
Particle transport in channels 
attracted recently much attention due to its importance in zeolites \cite{Keil2000}, 
biological\cite{Hille} and designed nanopores\cite{Pedone2010}, and other situations. 
This activity was stimulated by the interest to shape and size selective catalysis \cite{Cheng2008}, 
particles' separation\cite{Howorka2009,*Hanggi2011} and to polymer translocation \cite{Muthukumar2001,*Dekker2007}.
The progress in experiments boosted theoretical activities \cite{Reguera2006,*Burada2008}. 
Works by Jacobs and Zwanzig \cite{Zwanzig1992} lead to the so-called Fick-Jacobs (FJ)
approach, in which the elimination of transversal degrees of freedom (assuming fast
equilibration in transversal directions) results in an effective description for the longitudinal coordinate evolving 
in the potential of mean force. The approach found its application for biased transport through 
periodic planar \cite{Burada2008,Martens2011,*Martens2011b} 
and three-dimensional \cite{Dagdug2011} channels with smoothly varying cross-section. 
Other approaches were applied to transport through 
cylindrical septate channels \cite{Borromeo2010}, tubes formed by
spherical compartments \cite{Berezhkovskii2010b} and channels with abrupt changes in cross-section \cite{Kalinay2010,*Makhnovskii2010}. 

The applicability of the FJ approach depends on the existence of a hierarchy of relaxation times
governed by the geometry of the channel and by friction. This hierarchy guarantees
the time scale separation and the equipartition of energy both being necessary conditions for applying the method.
The core physical assumption behind the FJ approach is that the the dynamics of particles in a fluid inside the channel
is the overdamped Langevin one \cite{50yKramers}. 
In this work we study the impact of the friction coefficient $\gamma$ and of external bias $f$ on the transport in the channel and
concentrate on its influence on equipartition. 

\begin{figure}[t]
  \centering
  \includegraphics[width=0.99\linewidth]{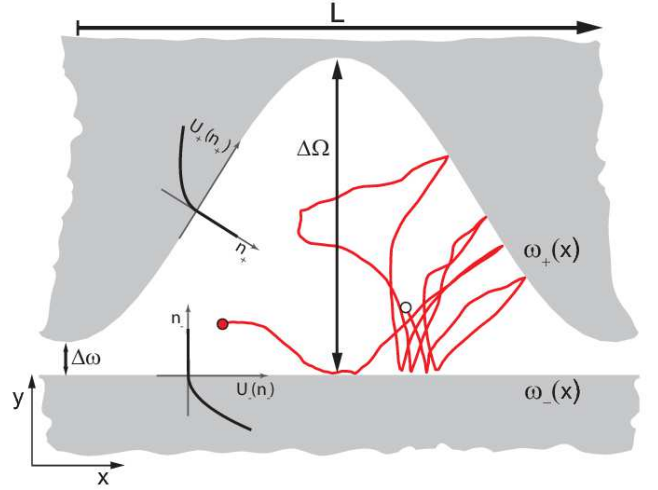}
  \caption{(Color online) Sketch of a segment of a sinusoidally-shaped  $2$D channel and  an example for a particle trajectory for $\gamma=1$ and $f=1$. The constant force $f$ is pointing in longitudinal ($x$-) direction. Shown are also the potentials $U_\pm(n_\pm)$ of the particle-wall interaction. 
  \label{fig:Fig1} }
\end{figure}

We consider Brownian particles with mass $m$ in a $2$D channel with top and bottom boundaries given by the functions 
$\omega_+(x)$ and $\omega_-(x)$ respectively, both periodic with period $L$, see \figref{fig:Fig1}. 
The particles are subject to an external static force $\V{f}$ acting along the longitudinal ($x$) 
direction (potential $U(x,y)=-f\,x$). The hydrodynamic interactions within the system 
can be neglected provided particles are small \cite{Fuchs} and their concentration is low. 
We define the system of units with $m = L = k_B T =1$. The unit of time in this system is 
$\tau=L\sqrt{m/k_BT}$. The velocity $\V{v}$ of the particle of particles is then governed by the Langevin equation
\begin{align}
 \V{\dot{v}}=\,-{\gamma}\,\V{v}-\nabla_\V{q} U(x,y)+\sqrt{2 \gamma}\,\mbox{\boldmath$\xi$}(t) \label{eq:eom}
\end{align}
with delta-correlated Gaussian random 
force $\mbox{\boldmath$\xi$}=(\xi_x,\xi_y)$:  $\av{\xi_i(t)} = 0$,
$\av{\xi_i(t) \xi_j(s)}=\delta_{ij}\delta(t-s)$; $i,j$ are $x$ or $y$. 

The evolution of the probability density function (PDF) $P\bracket{\V{q},\V{v},t}$ of position $\V{q}=(x,y)$ and velocity $\V{v}=(v_x,v_y)$ corresponding to \eq{eq:eom}
is governed by the Klein-Kramers equation
\cite{50yKramers} $\partial_t P=\,L_{x,v_x} P+L_{y,v_y} P\,,$ with $L_{q,v}=\,-v \partial_q+\partial_q U \partial_v +\gamma\partial_v \left[v+\partial_v\right]$                                                                                                                 
with no-flow condition at the boundaries. Approximations to this equation can give rise to effective theories concentrating on relevant $x$-coordinate and suppressing the irrelevant $y$-one. Let us first discuss necessary conditions for such an effective description.

According to the Bayes theorem, the joint PDF of the position and the 
velocity is given by the product
$ P(\V{q},\V{v},t) = \Phi(y,v_y|x,v_x,t)\,p(x,v_x,t)$ of the marginal probability density
$ p(x,v_x,t)=\,\int_{\omega_-(x)}^{\omega_+(x)}dy\int_{-\infty}^{\infty} dv_y\, P(\V{q},\V{v},t) $
and the joint PDF of $y$ and $v_y$ conditioned on $x,v_x$, and $t$. The fast relaxation approximation\cite{Berezhkovskii2011} assumes
that $\Phi(y,v_y|x,v_x,t)$ is equal to equilibrium PDF of $y$ and $v_y$, conditioned on $x$:
\begin{equation}
\Phi(y,v_y|x,v_x,t) = \Phi(y,v_y|x) 
\label{eq:prob_product}
\end{equation}
with $\Phi(y,v_y|x) =\,e^{-[v_y^2/2+U(x,y)]}\Big/\sqrt{2\pi}\int_{\omega_-(x)}^{\omega_+(x)} e^{-U(x,y)} dy$.
In this case the full dynamics, \eq{eq:eom}, can be replaced by the motion of a particle in the potential $A(x)$ of mean force
defined by
\begin{align}
 \p{A(x)}{x}=&\intl{\omega_-(x)}{\omega_+(x)}dy \intl{-\infty}{\infty} dv_y \partial_x U(x,y) \Phi(y,v_y|x)\,, \label{eq:def_meanforce}
\intertext{yielding}
 \dot{v}_x=\,&-\gamma\,v_x-\partial_x A(x)+\sqrt{2\gamma}\,\xi_x(t)\,. \label{eq:eom_meanforce}
\end{align}
The difference between the external force and the mean force,
$\delta F_x(y)=-\partial_x U(x,y)+\partial_xA(x)$, results in an additional effective deterministic force as well as in position-dependent effective friction \cite{Berezhkovskii2011} and diffusion coefficients \cite{Martens2011}. 

The assumption \eq{eq:prob_product} is valid if (i) the distribution of $y$ 
relaxes fast enough to the equilibrium one, 
(ii) equipartition of the kinetic energies corresponding to $v_x$ and $v_y$ holds, 
and (iii) the two velocity components are uncorrelated at any time. 
Burada et al. \cite{Burada2007} analyzed time scales involved in the problem. 
These are the times  $\tau_y=\gamma\,\Delta y^2/2$ and $\tau_x=\gamma\,\Delta x^2/2$ to diffuse over distances $\Delta y$ and $\Delta x$, respectively, 
the characteristic times associated with the 
drift $\tau_\mathrm{drift}^x=\mathrm{min}(\gamma\,\Delta x/f,\gamma\,\Delta x/v_x)$ and $\tau_\mathrm{drift}^y=\gamma\,\Delta y/v_y$, and 
the velocity correlation time $\tau_\mathrm{cor}=1/\gamma$. 
A general criterion that has to be satisfied is that 
$\mathrm{max}\bracket{\tau_y/\tau_x,\tau_\mathrm{drift}^y/\tau_\mathrm{drift}^x,\tau_\mathrm{cor}/\tau_\mathrm{drift}^y}\ll 1$.
This can be achieved either for strongly anisotropic friction $\gamma_x\gg\gamma_y$ \cite{Berezhkovskii2011,Kalinay2006} or for relatively flat boundaries \cite{Martens2011,*Martens2011b}.

In the high friction limit, $\gamma\gg 1$, the Klein-Kramers equation associated with \eq{eq:eom_meanforce} 
simplifies to the FJ equation \cite{Zwanzig1992} $\partial_t p(x,t)=\partial_x\left[e^{-A(x)}\partial_x\bracket{e^{A(x)} p(x,t)}\right]$,
and the analytic expression for the potential of mean force $A(x)$ can be derived along the lines of Ref.\cite{Sokolov2010}.
To do this, we mimic the interaction of the particles with walls by a quadratic potential growing in the direction normal to the wall, $U_\pm(n_\pm)=\,\frac{\kappa}{2}n_\pm^2$
with interaction strength $\kappa$ and $n_\pm$ being the coordinate along the normal to the upper or lower boundary taken at the point
$(x, \omega_\pm(x))$. 
For $x$ fixed, the energy depends only on $y$ and is given by $U_\pm(x,y)=0.5\kappa\bracket{y-\omega_\pm(x)}^2 \cos^2 \alpha_\pm$ with $\alpha_\pm=\arctan\bracket{d\omega_\pm(x)/dx}$ for $y<\omega_-(x)$ and $y>\omega_+(x)$, and vanishes otherwise. \eq{eq:def_meanforce} then reduces to $\partial_x A(x)=\int_{-\infty}^\infty dy \int_{-\infty}^\infty dv_y  \bracket{\partial_xU+\partial_xU_\pm} \Phi(y,v_y|x)$. 
Integrating over $v_y$ and interchanging the derivative and integration yield $\partial_x A(x)=-\partial_x \ln\left[\int_{-\infty}^\infty dy \exp\bracket{-U-U_\pm}\right]$. 
The domain of integration over $y$ can be divided into three intervals $-\infty<y\leq \omega_-(x)$, $\omega_-(x)<y<\omega_+(x)$, and $\omega_+(x)<y<\infty$. 
Integral $\int_{-\infty}^{\omega_-(x)} dy \ldots$ can be evaluated explicitly by integrating by parts. The corresponding expression vanishes in the limit of hard walls, i.e. $\kappa \to \infty$. The same happens with the integral $\int_{\omega_+(x)} ^{\infty}dy \ldots\,$.
Consequently, the mean force reads
\begin{align}
 -\p{A(x)}{x}=\,\partial_x \ln\left[\intl{\omega_-(x)}{\omega_+(x)} dy \exp\bracket{-U(x,y)}\right]\,.
\end{align}
The potential of mean force is the free energy associated with the partition function $Z(x)=\int_{\omega_-(x)}^{\omega_+(x)} dy \exp\bracket{-U(x,y)}$ and 
does not depend on $\gamma$. This result relies on Maxwell distribution of $v_y$, i.e. on equipartition. As we proceed to show, this property breaks down if the motion is not overdamped.

In the following, we study the mobility $\mu/\mu_0=\lim_{t\to \infty} \gamma \av{x(t)}/(f\,t)$ and the effective diffusion coefficient $D_\mathrm{eff}/D_0=\lim_{t\to \infty} \gamma\,(\av{x(t)^2}-\av{x(t)}^2)/(2\,t)$ of particles moving in a sinusoidally-shaped \cite{Burada2008} channel with the top boundary given by
\begin{align}
 \omega_+\bracket{x}=&\,\frac{1}{2}\left[\Delta\Omega+\Delta\omega-\bracket{\Delta\Omega-\Delta\omega}\cos\bracket{2\pi\,x}\right]\,, \label{eq:conf}
\end{align}
and with flat bottom boundary $\omega_-(x)=0$,\figref{fig:Fig1}. $\Delta \Omega$ and $\Delta \omega$ denote the maximal and the minimal width of the channel, respectively, with aspect ratio $\delta=\Delta \omega/\Delta \Omega$.

\begin{figure}[h!]
 \includegraphics[width=0.92\linewidth]{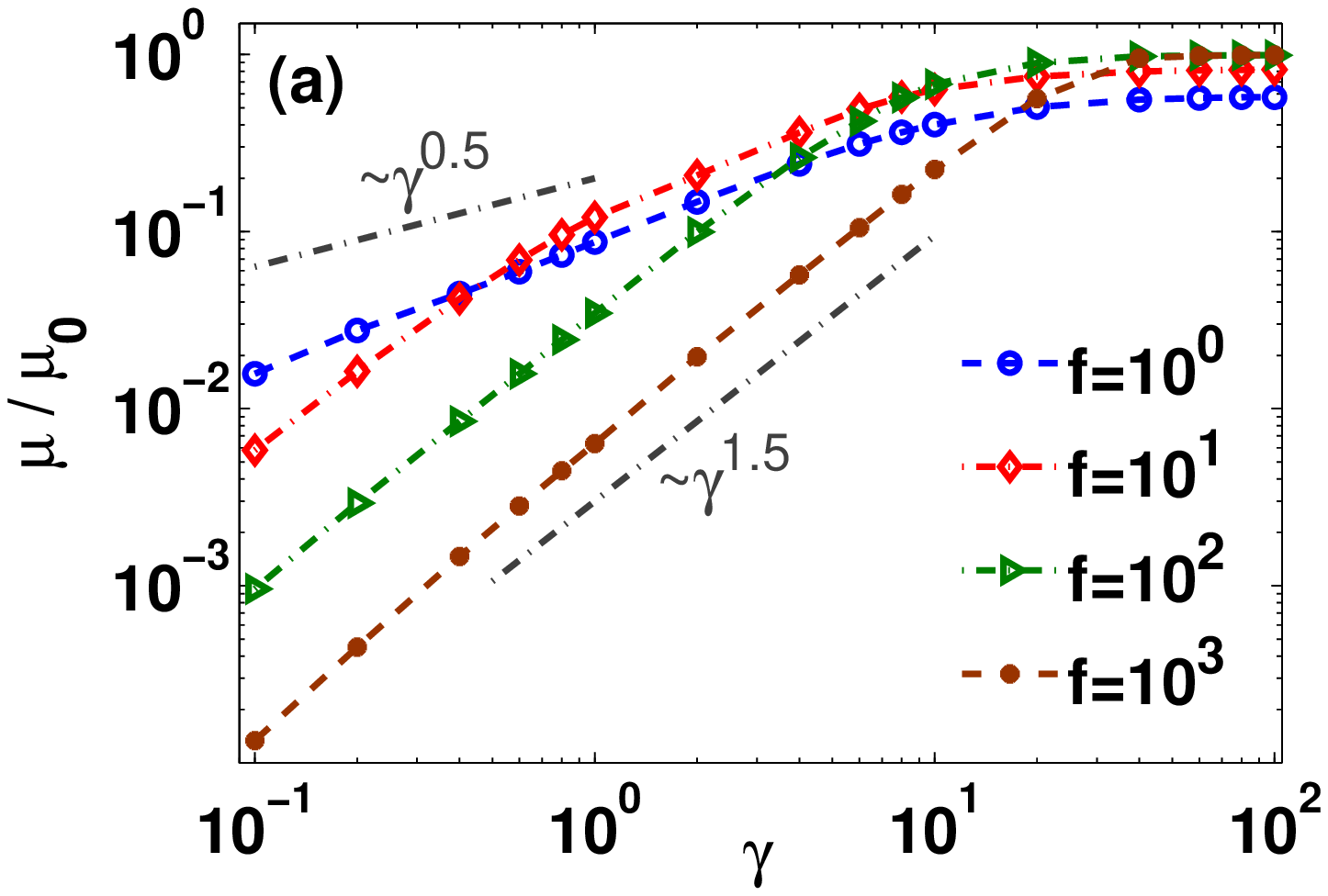}\\
 \includegraphics[width=0.92\linewidth]{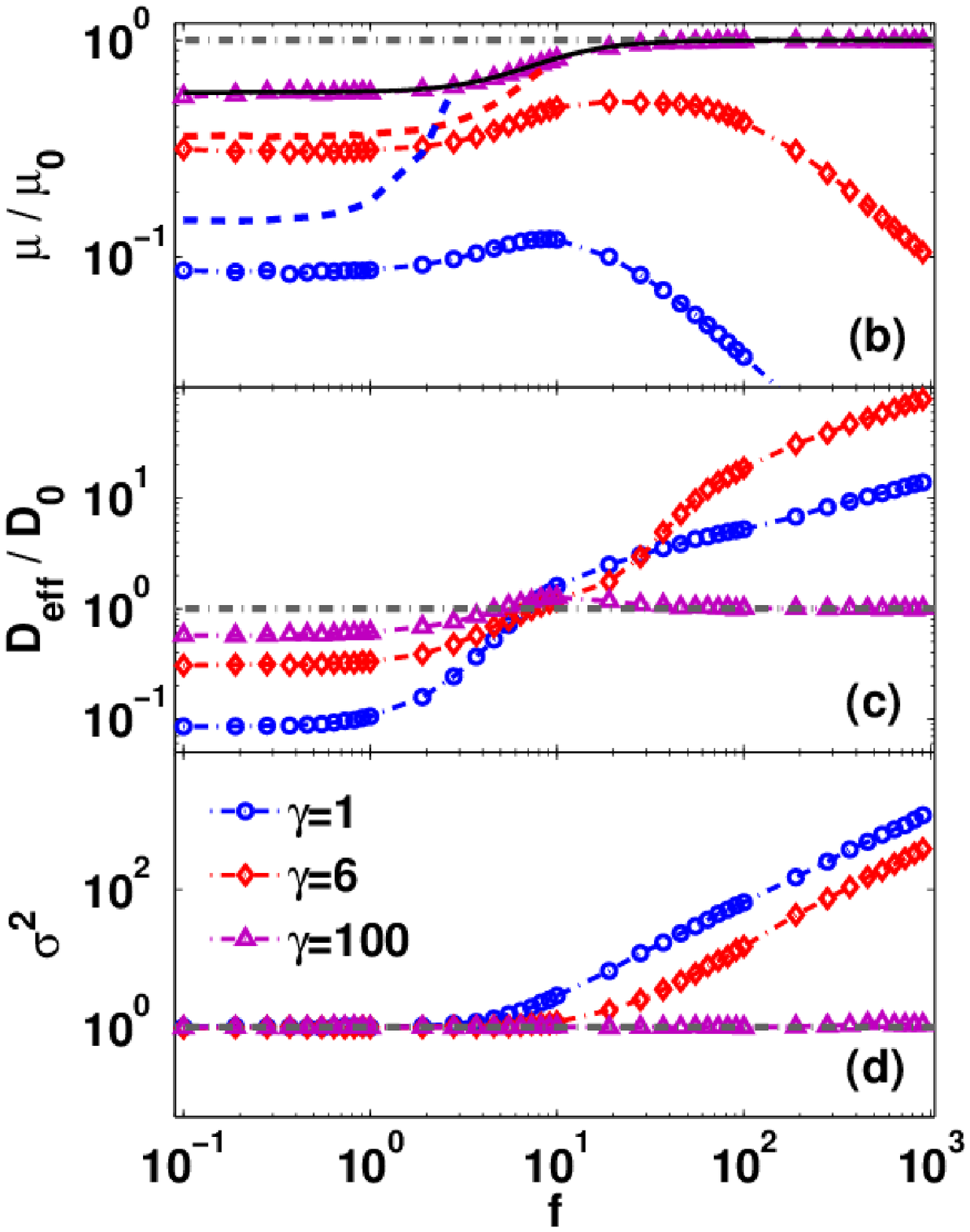}
\caption{(Color online) Results of simulation of full dynamics, \eq{eq:eom}, in a channel geometry of \figref{fig:Fig1} with  
$\Delta\omega=0.005$ and $\Delta\Omega=0.05$. Panel (a): Particle mobility $\mu/\mu_0$ 
as a function of $\gamma$ for different force magnitudes $f$. 
Panels (b) - (d): the force dependence of different dynamical characteristics of the system for  $\gamma=1,6$ and $100$. 
(b) The particle mobility, symbols. Superimposed are numerical results for reduced dynamics, \eq{eq:eom_meanforce} (dashed lines), 
and the analytical result for the overdamped case \eq{eq:mob_highfriction}, (solid line). 
(c) The effective diffusion coefficient, $D_\mathrm{eff}/D_0$. 
(d) The $2$nd central moment $\sigma^2$ of $v_y$. The horizontal dash-dotted lines indicate unity.} \label{fig:Fig2}
\end{figure}

\figref{fig:Fig2} shows the influence of the external force magnitude $f$ on the particle mobility (see panel (b)) for various friction coefficients $\gamma$. For $\gamma \gg 1$, one observes the known dependence of $\mu/\mu_0$ on $f$ \cite{Burada2008}. Starting from the asymptotic value $\mu/\mu_0=2\sqrt{\delta}/(1+\delta)$ for $f\ll 1$ (see Eq.~(45) in Ref.~\cite{Martens2011}), the mobility increases with force magnitude till the asymptotic value $\mu/\mu_0=1$ is reached for $f\to \infty$. 
It can be calculated using the Stratonovich formula \cite{50yKramers} giving rise to
\begin{equation}
\frac{\mu}{\mu_0}=\frac{f^2+\bracket{2\pi}^2}{f^2+\frac{\bracket{2\pi}^2}{2}\bracket{\sqrt{\delta}+1/\sqrt{\delta}}}. \label{eq:mob_highfriction}
\end{equation}
The result \eq{eq:mob_highfriction} matches perfectly the numerics for $\gamma=100$, see \figref{fig:Fig2}(b).

In the diffusion dominated regime, $f\leq 1$ in Figs.~\ref{fig:Fig2}(a)-(b), the mobility decreases with decreasing $\gamma$. 
Such a dependence is known for arbitrarily damped Brownian motion in periodic potentials \cite{50yKramers} and might witness for the applicability of 
the reduced description. For chosen geometry $\mu/\mu_0$ goes as $\simeq \gamma^{0.5}$; the exponent depends both on $\Delta\Omega$ and on $\Delta\omega$ (not shown). 
Moreover, the mobility and the effective diffusion coefficient coincide for $f\ll 1$ (see \figref{fig:Fig2} (c)) thus corroborating the Sutherland-Einstein relation \cite{Burada2008}. 

For larger forces the mobility increases until it reaches its maximum at $f_\mathrm{max}$ (depending on friction) and then decays 
as $\mu/\mu_0 \propto f^{-\alpha}$ with $\alpha<1$. 
The particle mobility for the reduced dynamics in the presence of the potential of mean force $A(x)$ obtained by simulating \eq{eq:eom_meanforce}
is shown in \figref{fig:Fig2} (b) by dashed lines. The approximation overestimates the true mobility for all $f$ but is sufficiently accurate for 
$\gamma \geq 5$ and $f< f_\mathrm{max}$. For $\gamma=1$ the discrepancy is large even for $f\ll1$.
Introducing the position-dependent friction coefficient 
$\gamma(x)$, as proposed in Ref.\cite{Berezhkovskii2011}, gives corrections of the order of  $\bracket{\Delta\Omega}^2$ and 
does not improve the agreement.

\begin{figure}
  \centering
  \includegraphics[width=0.95\linewidth]{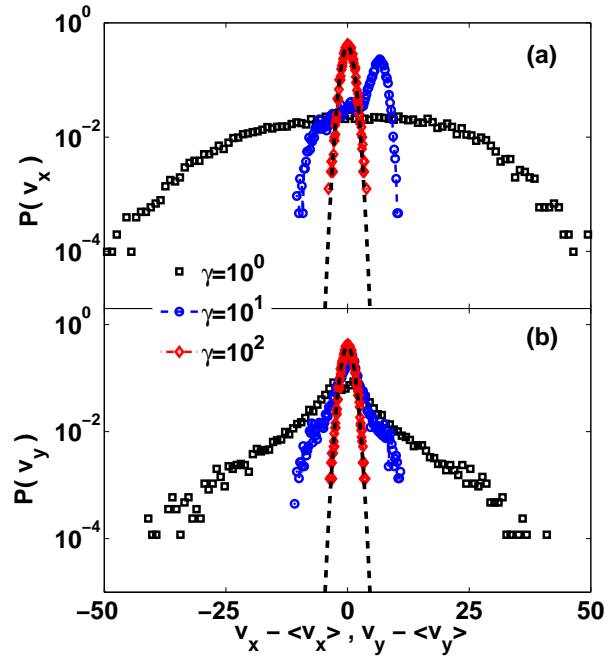}
  \caption{(Color online) Normalized probability distribution functions of $v_x$ and $v_y$ as functions of the friction coefficient $\gamma$ for $f=100$. The mean values are $\av{v_x}\approx 3.46,6.78,0.99$ (for $\gamma=10^0,10^1,10^2$) and $\av{v_y}\simeq 0$ (for all $\gamma$ values). The black dashed lines indicate the Maxwell velocity distribution $\exp\bracket{-v_{x,y}^2/2}/\sqrt{2\pi}$.}
  \label{fig:Fig3}
\end{figure}
Our derivation of the effective dynamics
implied the Maxwell distribution of $v_y$ and the homogeneous distribution of $y$. 
In \figref{fig:Fig3} we present the velocity distributions $P(v_x)$ and $P(v_y)$ centered at their 
means $\av{v_x}$ and $\av{v_y}$ for different $\gamma$ and for fixed $f=100$. 
The distribution $P(v_x)$,  \figref{fig:Fig3} (a), undergoes a transition from a  normal (Maxwell) distribution with variance $1$ for $\gamma= 100$ 
over a broader bimodal distribution for $\gamma=10$ to the broad symmetric function for $\gamma=1$. 
In contrast, $P(v_y)$,  \figref{fig:Fig3} (b), stays symmetric regardless of the value of $\gamma$. Similar to $P(v_x)$, the smaller the friction 
the broader the distribution. 
The deviation of $P(v_y)$ from the Maxwell distribution is given by its second central moment $\sigma^2=\av{v_y^2}-\av{v_y}^2$, \figref{fig:Fig2} (d),
which is unity (in our units) for the Maxwell one. 
The distribution of $v_y$ is Maxwellian independent of $f$ for sufficiently high friction  
for any $f$. For smaller friction the transversal velocity distribution broadens, $\sigma^2 \propto f^{\beta}$ with $\beta>1$, 
if $f$ exceeds a critical magnitude $f_c$ which practically coincides with  $f_\mathrm{max}$. 
In other words, the decrease of the particle mobility goes hand in hand with violation of equipartition. The $2$nd central moment of $v_x$ shows the same dependence on the external force magnitude and on the friction coefficient as the one of $v_y$.

Finally, we discuss the impact of $\gamma$ on the effective diffusion coefficient. In the high friction limit we reproduce the results of Ref.\cite{Burada2008}:
starting from the value of $D_\mathrm{eff}/D_0=2\sqrt{\delta}/(1+\delta)$ for $f \to 0$ 
the effective diffusivity grows with increasing $f$ until it reaches its maximum and then decays and finally tends to $D_\mathrm{eff}/D_0=1$ for $f\to \infty$. 
For $f\ll 1$ the effective diffusivity grows with friction. 
For sufficiently strong forces, $f>10$, the behavior of $D_\mathrm{eff}/D_0$ as a function of $\gamma$ is non-monotonic,
passing through a maximum at some value of $\gamma$ which depends on the force magnitude and on the channel's geometry. 
\begin{figure}
  \centering
  \includegraphics[width=0.95\linewidth]{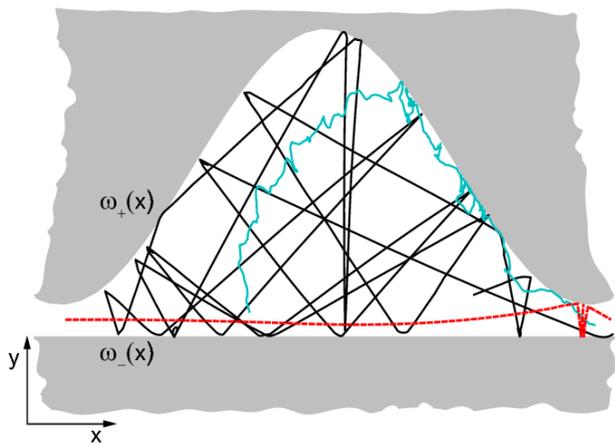}
  \caption{(Color online) Particle trajectories for $\gamma=1$ (black solid line and dashed line) and for $\gamma=100$ (bright solid line). The force magnitude is $f=100$.
The geometry parameters are $\Delta\Omega=1$ and $\Delta\omega=0.1$.}
  \label{fig:Fig_traj}
\end{figure}

Let us discuss the nature of the equipartition violation.
In free motion the velocities ''forget'' their initial values and assume equilibrium distributions 
for $t>1/\gamma$ for any $\gamma$. The value of $\gamma$ influences however the properties of confined motion, see \figref{fig:Fig_traj},
changing from erratic for high friction to almost regular - known for the deterministic case \cite{Gaspard2002,*Cecconi2003} - for $\gamma \to 0$.
Due to the reflection at the boundaries
the acceleration caused by the external force $f$ is transferred from the longitudinal to the transversal velocity component violating equipartition
and leading to the monotonous growth of $\sigma^2$ with $f$. 
The behavior in the $x$-direction is more complex. On one hand, the particles reflected at an ''optimal'' angle can
fly over several cells to the left or to the right. 
We found that the probability of long stretches increases with force magnitude and with decreasing friction,
and that long excursions into the direction of the force are always more probable than in the opposite one.
On the other hand, some particles undergo repeated collisions with walls where these are almost parallel, i.e. close to the minima and maxima
of the channel's width, see dashed line in \figref{fig:Fig_traj}, leading to trapping. The interplay of both effects leads to complex behavior and non-monotonicities described above.

\textbf{Conclusions} 
Let us summarize our findings. We investigated the impact of the friction coefficient on biased Brownian motion of point-like particles in a two-dimensional channel with smoothly varying width. For high friction, the adiabatic elimination of the transverse degrees of freedom results to an effective description for the slow $x$-coordinate involving the potential of mean force $A(x)$, leading to an exact analytical result for the particle mobility \eq{eq:mob_highfriction} valid for narrow channels \cite{Martens2011}. Comparing the results for
reduced dynamics with numerical results for the full problem, we show that the reduced description overestimates the mobility, although the accuracy of the approximation 
is sufficiently good for $\gamma \geq 5$ and small forces $f< f_\mathrm{max}$. There exists a characteristic force strength $f_c$ beyond which the 
reduced description fails. The force $f_c$ become less with decreasing friction. The origin of the failure of the effective description is the violation
of equipartition for the fast coordinate $y$ and velocity $v_y$. 

The authors are thankful to P. H\"{a}nggi and G. Schmid for useful discussions and acknowledge financial support by the VW Foundation via project I/83903.


\begin{thebibliography}{26}%
\makeatletter
\providecommand \@ifxundefined [1]{%
 \@ifx{#1\undefined}
}%
\providecommand \@ifnum [1]{%
 \ifnum #1\expandafter \@firstoftwo
 \else \expandafter \@secondoftwo
 \fi
}%
\providecommand \@ifx [1]{%
 \ifx #1\expandafter \@firstoftwo
 \else \expandafter \@secondoftwo
 \fi
}%
\providecommand \natexlab [1]{#1}%
\providecommand \enquote  [1]{``#1''}%
\providecommand \bibnamefont  [1]{#1}%
\providecommand \bibfnamefont [1]{#1}%
\providecommand \citenamefont [1]{#1}%
\providecommand \href@noop [0]{\@secondoftwo}%
\providecommand \href [0]{\begingroup \@sanitize@url \@href}%
\providecommand \@href[1]{\@@startlink{#1}\@@href}%
\providecommand \@@href[1]{\endgroup#1\@@endlink}%
\providecommand \@sanitize@url [0]{\catcode `\\12\catcode `\$12\catcode
  `\&12\catcode `\#12\catcode `\^12\catcode `\_12\catcode `\%12\relax}%
\providecommand \@@startlink[1]{}%
\providecommand \@@endlink[0]{}%
\providecommand \url  [0]{\begingroup\@sanitize@url \@url }%
\providecommand \@url [1]{\endgroup\@href {#1}{\urlprefix }}%
\providecommand \urlprefix  [0]{URL }%
\providecommand \Eprint [0]{\href }%
\providecommand \doibase [0]{http://dx.doi.org/}%
\providecommand \selectlanguage [0]{\@gobble}%
\providecommand \bibinfo  [0]{\@secondoftwo}%
\providecommand \bibfield  [0]{\@secondoftwo}%
\providecommand \translation [1]{[#1]}%
\providecommand \BibitemOpen [0]{}%
\providecommand \bibitemStop [0]{}%
\providecommand \bibitemNoStop [0]{.\EOS\space}%
\providecommand \EOS [0]{\spacefactor3000\relax}%
\providecommand \BibitemShut  [1]{\csname bibitem#1\endcsname}%
\let\auto@bib@innerbib\@empty
\bibitem [{\citenamefont {Keil}, \citenamefont {Krishna},\ and\ \citenamefont
  {Coppens}(2000)}]{Keil2000}%
  \BibitemOpen
  \bibfield  {author} {\bibinfo {author} {\bibfnamefont {F.}~\bibnamefont
  {Keil}}, \bibinfo {author} {\bibfnamefont {R.}~\bibnamefont {Krishna}}, \
  and\ \bibinfo {author} {\bibfnamefont {M.}~\bibnamefont {Coppens}},\
  }\href@noop {} {\bibfield  {journal} {\bibinfo  {journal} {Rev. Chem. Eng.}\
  }\textbf {\bibinfo {volume} {16}},\ \bibinfo {pages} {71} (\bibinfo {year}
  {2000})}\BibitemShut {NoStop}%
\bibitem [{\citenamefont {Hille}(2001)}]{Hille}%
  \BibitemOpen
  \bibfield  {author} {\bibinfo {author} {\bibfnamefont {B.}~\bibnamefont
  {Hille}},\ }\href@noop {} {\emph {\bibinfo {title} {Ion Channels of Excitable
  Membranes}}},\ \bibinfo {edition} {3rd}\ ed.\ (\bibinfo  {publisher} {Sinauer
  Associates},\ \bibinfo {year} {2001})\BibitemShut {NoStop}%
\bibitem [{\citenamefont {Pedone}\ \emph {et~al.}(2010)\citenamefont {Pedone},
  \citenamefont {Langecker}, \citenamefont {Muenzer}, \citenamefont {Wei},
  \citenamefont {Nagel},\ and\ \citenamefont {Rant}}]{Pedone2010}%
  \BibitemOpen
  \bibfield  {author} {\bibinfo {author} {\bibfnamefont {D.}~\bibnamefont
  {Pedone}}, \bibinfo {author} {\bibfnamefont {M.}~\bibnamefont {Langecker}},
  \bibinfo {author} {\bibfnamefont {A.~M.}\ \bibnamefont {Muenzer}}, \bibinfo
  {author} {\bibfnamefont {R.}~\bibnamefont {Wei}}, \bibinfo {author}
  {\bibfnamefont {R.~D.}\ \bibnamefont {Nagel}}, \ and\ \bibinfo {author}
  {\bibfnamefont {U.}~\bibnamefont {Rant}},\ }\href@noop {} {\bibfield
  {journal} {\bibinfo  {journal} {J. Phys. Condens. Matter}\ }\textbf {\bibinfo
  {volume} {22}},\ \bibinfo {pages} {454115} (\bibinfo {year}
  {2010})}\BibitemShut {NoStop}%
\bibitem [{\citenamefont {Cheng}, \citenamefont {Sheng},\ and\ \citenamefont
  {Tsao}(2008)}]{Cheng2008}%
  \BibitemOpen
  \bibfield  {author} {\bibinfo {author} {\bibfnamefont {K.~L.}\ \bibnamefont
  {Cheng}}, \bibinfo {author} {\bibfnamefont {Y.~J.}\ \bibnamefont {Sheng}}, \
  and\ \bibinfo {author} {\bibfnamefont {H.~K.}\ \bibnamefont {Tsao}},\ }\href
  {\doibase DOI:10.1063/1.3009621} {\bibfield  {journal} {\bibinfo  {journal}
  {J. Chem. Phys.}\ }\textbf {\bibinfo {volume} {129}},\ \bibinfo {pages}
  {184901} (\bibinfo {year} {2008})}\BibitemShut {NoStop}%
\bibitem [{\citenamefont {Howorka}\ and\ \citenamefont
  {Siwy}(2009)}]{Howorka2009}%
  \BibitemOpen
  \bibfield  {author} {\bibinfo {author} {\bibfnamefont {S.}~\bibnamefont
  {Howorka}}\ and\ \bibinfo {author} {\bibfnamefont {Z.}~\bibnamefont {Siwy}},\
  }\href@noop {} {\bibfield  {journal} {\bibinfo  {journal} {Chem. Soc. Rev.}\
  }\textbf {\bibinfo {volume} {38}},\ \bibinfo {pages} {2360} (\bibinfo {year}
  {2009})}\BibitemShut {NoStop}%
\bibitem [{\citenamefont {Reguera}\ \emph {et~al.}(2012)\citenamefont
  {Reguera}, \citenamefont {Luque}, \citenamefont {Burada}, \citenamefont
  {Schmid}, \citenamefont {Rub\'\i},\ and\ \citenamefont
  {H\"anggi}}]{Hanggi2011}%
  \BibitemOpen
  \bibfield  {author} {\bibinfo {author} {\bibfnamefont {D.}~\bibnamefont
  {Reguera}}, \bibinfo {author} {\bibfnamefont {A.}~\bibnamefont {Luque}},
  \bibinfo {author} {\bibfnamefont {P.~S.}\ \bibnamefont {Burada}}, \bibinfo
  {author} {\bibfnamefont {G.}~\bibnamefont {Schmid}}, \bibinfo {author}
  {\bibfnamefont {J.~M.}\ \bibnamefont {Rub\'\i}}, \ and\ \bibinfo {author}
  {\bibfnamefont {P.}~\bibnamefont {H\"anggi}},\ }\href {\doibase
  10.1103/PhysRevLett.108.020604} {\bibfield  {journal} {\bibinfo  {journal}
  {Phys. Rev. Lett.}\ }\textbf {\bibinfo {volume} {108}},\ \bibinfo {pages}
  {020604} (\bibinfo {year} {2012})}\BibitemShut {NoStop}%
\bibitem [{\citenamefont {Muthukumar}(2001)}]{Muthukumar2001}%
  \BibitemOpen
  \bibfield  {author} {\bibinfo {author} {\bibfnamefont {M.}~\bibnamefont
  {Muthukumar}},\ }\href {\doibase 10.1103/PhysRevLett.86.3188} {\bibfield
  {journal} {\bibinfo  {journal} {Phys. Rev. Lett.}\ }\textbf {\bibinfo
  {volume} {86}},\ \bibinfo {pages} {3188} (\bibinfo {year}
  {2001})}\BibitemShut {NoStop}%
\bibitem [{\citenamefont {Dekker}(2007)}]{Dekker2007}%
  \BibitemOpen
  \bibfield  {author} {\bibinfo {author} {\bibfnamefont {C.}~\bibnamefont
  {Dekker}},\ }\href@noop {} {\bibfield  {journal} {\bibinfo  {journal} {Nature
  Nanotech.}\ }\textbf {\bibinfo {volume} {2}},\ \bibinfo {pages} {209}
  (\bibinfo {year} {2007})}\BibitemShut {NoStop}%
\bibitem [{\citenamefont {Reguera}\ \emph {et~al.}(2006)\citenamefont
  {Reguera}, \citenamefont {Schmid}, \citenamefont {Burada}, \citenamefont
  {Rub\'\i}, \citenamefont {Reimann},\ and\ \citenamefont
  {H{\"a}nggi}}]{Reguera2006}%
  \BibitemOpen
  \bibfield  {author} {\bibinfo {author} {\bibfnamefont {D.}~\bibnamefont
  {Reguera}}, \bibinfo {author} {\bibfnamefont {G.}~\bibnamefont {Schmid}},
  \bibinfo {author} {\bibfnamefont {P.~S.}\ \bibnamefont {Burada}}, \bibinfo
  {author} {\bibfnamefont {J.~M.}\ \bibnamefont {Rub\'\i}}, \bibinfo {author}
  {\bibfnamefont {P.}~\bibnamefont {Reimann}}, \ and\ \bibinfo {author}
  {\bibfnamefont {P.}~\bibnamefont {H{\"a}nggi}},\ }\href@noop {} {\bibfield
  {journal} {\bibinfo  {journal} {Phys. Rev. Lett.}\ }\textbf {\bibinfo
  {volume} {96}},\ \bibinfo {pages} {130603} (\bibinfo {year}
  {2006})}\BibitemShut {NoStop}%
\bibitem [{\citenamefont {Burada}\ \emph {et~al.}(2008)\citenamefont {Burada},
  \citenamefont {Schmid}, \citenamefont {Talkner}, \citenamefont {H{\"a}nggi},
  \citenamefont {Reguera},\ and\ \citenamefont {Rub\'\i}}]{Burada2008}%
  \BibitemOpen
  \bibfield  {author} {\bibinfo {author} {\bibfnamefont {P.~S.}\ \bibnamefont
  {Burada}}, \bibinfo {author} {\bibfnamefont {G.}~\bibnamefont {Schmid}},
  \bibinfo {author} {\bibfnamefont {P.}~\bibnamefont {Talkner}}, \bibinfo
  {author} {\bibfnamefont {P.}~\bibnamefont {H{\"a}nggi}}, \bibinfo {author}
  {\bibfnamefont {D.}~\bibnamefont {Reguera}}, \ and\ \bibinfo {author}
  {\bibfnamefont {J.~M.}\ \bibnamefont {Rub\'\i}},\ }\href@noop {} {\bibfield
  {journal} {\bibinfo  {journal} {BioSystems}\ }\textbf {\bibinfo {volume}
  {93}},\ \bibinfo {pages} {16} (\bibinfo {year} {2008})}\BibitemShut {NoStop}%
\bibitem [{\citenamefont {Zwanzig}(1992)}]{Zwanzig1992}%
  \BibitemOpen
  \bibfield  {author} {\bibinfo {author} {\bibfnamefont {R.}~\bibnamefont
  {Zwanzig}},\ }\href {\doibase 10.1021/j100189a004} {\bibfield  {journal}
  {\bibinfo  {journal} {J. Phys. Chem.}\ ,\ \bibinfo {pages} {3926}} (\bibinfo
  {year} {1992})}\BibitemShut {NoStop}%
\bibitem [{\citenamefont {Martens}\ \emph
  {et~al.}(2011{\natexlab{a}})\citenamefont {Martens}, \citenamefont {Schmid},
  \citenamefont {Schimansky-Geier},\ and\ \citenamefont
  {H{\"a}nggi}}]{Martens2011}%
  \BibitemOpen
  \bibfield  {author} {\bibinfo {author} {\bibfnamefont {S.}~\bibnamefont
  {Martens}}, \bibinfo {author} {\bibfnamefont {G.}~\bibnamefont {Schmid}},
  \bibinfo {author} {\bibfnamefont {L.}~\bibnamefont {Schimansky-Geier}}, \
  and\ \bibinfo {author} {\bibfnamefont {P.}~\bibnamefont {H{\"a}nggi}},\
  }\href {\doibase 10.1103/PhysRevE.83.051135} {\bibfield  {journal} {\bibinfo
  {journal} {Phys. Rev. E}\ }\textbf {\bibinfo {volume} {83}},\ \bibinfo
  {pages} {051135} (\bibinfo {year} {2011}{\natexlab{a}})}\BibitemShut
  {NoStop}%
\bibitem [{\citenamefont {Martens}\ \emph
  {et~al.}(2011{\natexlab{b}})\citenamefont {Martens}, \citenamefont {Schmid},
  \citenamefont {Schimansky-Geier},\ and\ \citenamefont
  {Hanggi}}]{Martens2011b}%
  \BibitemOpen
  \bibfield  {author} {\bibinfo {author} {\bibfnamefont {S.}~\bibnamefont
  {Martens}}, \bibinfo {author} {\bibfnamefont {G.}~\bibnamefont {Schmid}},
  \bibinfo {author} {\bibfnamefont {L.}~\bibnamefont {Schimansky-Geier}}, \
  and\ \bibinfo {author} {\bibfnamefont {P.}~\bibnamefont {H{\"a}nggi}},\ }\href
  {\doibase 10.1063/1.3658621} {\bibfield  {journal} {\bibinfo  {journal}
  {Chaos}\ }\textbf {\bibinfo {volume} {21}},\ \bibinfo {eid} {047518}
  (\bibinfo {year} {2011}{\natexlab{b}})}\BibitemShut {NoStop}%
\bibitem [{\citenamefont {Dagdug}\ \emph {et~al.}(2011)\citenamefont {Dagdug},
  \citenamefont {Berezhkovskii}, \citenamefont {Makhnovskii}, \citenamefont
  {Zitserman},\ and\ \citenamefont {Bezrukov}}]{Dagdug2011}%
  \BibitemOpen
  \bibfield  {author} {\bibinfo {author} {\bibfnamefont {L.}~\bibnamefont
  {Dagdug}}, \bibinfo {author} {\bibfnamefont {A.~M.}\ \bibnamefont
  {Berezhkovskii}}, \bibinfo {author} {\bibfnamefont {Y.~A.}\ \bibnamefont
  {Makhnovskii}}, \bibinfo {author} {\bibfnamefont {V.~Y.}\ \bibnamefont
  {Zitserman}}, \ and\ \bibinfo {author} {\bibfnamefont {S.~M.}\ \bibnamefont
  {Bezrukov}},\ }\href {\doibase 10.1063/1.3561680} {\bibfield  {journal}
  {\bibinfo  {journal} {J. Chem. Phys.}\ }\textbf {\bibinfo {volume} {134}},\
  \bibinfo {pages} {101102} (\bibinfo {year} {2011})}\BibitemShut {NoStop}%
\bibitem [{\citenamefont {Borromeo}\ and\ \citenamefont
  {Marchesoni}(2010)}]{Borromeo2010}%
  \BibitemOpen
  \bibfield  {author} {\bibinfo {author} {\bibfnamefont {M.}~\bibnamefont
  {Borromeo}}\ and\ \bibinfo {author} {\bibfnamefont {F.}~\bibnamefont
  {Marchesoni}},\ }\href@noop {} {\bibfield  {journal} {\bibinfo  {journal}
  {Chem. Phys.}\ }\textbf {\bibinfo {volume} {375}},\ \bibinfo {pages} {536}
  (\bibinfo {year} {2010})}\BibitemShut {NoStop}%
\bibitem [{\citenamefont {Berezhkovskii}\ \emph {et~al.}(2010)\citenamefont
  {Berezhkovskii}, \citenamefont {Dagdug}, \citenamefont {Makhnovskii},\ and\
  \citenamefont {Zitserman}}]{Berezhkovskii2010b}%
  \BibitemOpen
  \bibfield  {author} {\bibinfo {author} {\bibfnamefont {A.~M.}\ \bibnamefont
  {Berezhkovskii}}, \bibinfo {author} {\bibfnamefont {L.}~\bibnamefont
  {Dagdug}}, \bibinfo {author} {\bibfnamefont {Y.~A.}\ \bibnamefont
  {Makhnovskii}}, \ and\ \bibinfo {author} {\bibfnamefont {V.~Y.}\ \bibnamefont
  {Zitserman}},\ }\href@noop {} {\bibfield  {journal} {\bibinfo  {journal} {J.
  Chem. Phys.}\ }\textbf {\bibinfo {volume} {132}},\ \bibinfo {pages} {221104}
  (\bibinfo {year} {2010})}\BibitemShut {NoStop}%
\bibitem [{\citenamefont {Kalinay}\ and\ \citenamefont
  {Percus}(2010)}]{Kalinay2010}%
  \BibitemOpen
  \bibfield  {author} {\bibinfo {author} {\bibfnamefont {P.}~\bibnamefont
  {Kalinay}}\ and\ \bibinfo {author} {\bibfnamefont {J.~K.}\ \bibnamefont
  {Percus}},\ }\href {\doibase 10.1103/PhysRevE.82.031143} {\bibfield
  {journal} {\bibinfo  {journal} {Phys. Rev. E}\ }\textbf {\bibinfo {volume}
  {82}},\ \bibinfo {pages} {031143} (\bibinfo {year} {2010})}\BibitemShut
  {NoStop}%
\bibitem [{\citenamefont {Makhnovskii}, \citenamefont {Berezhkovskii},\ and\
  \citenamefont {Zitserman}(2010)}]{Makhnovskii2010}%
  \BibitemOpen
  \bibfield  {author} {\bibinfo {author} {\bibfnamefont {Y.~A.}~\bibnamefont
  {Makhnovskii}}, \bibinfo {author} {\bibfnamefont {A.~M.}\ \bibnamefont
  {Berezhkovskii}}, \ and\ \bibinfo {author} {\bibfnamefont {V.~Y.}\
  \bibnamefont {Zitserman}},\ }\href {\doibase DOI:
  10.1016/j.chemphys.2010.04.012} {\bibfield  {journal} {\bibinfo  {journal}
  {Chem. Phys}\ }\textbf {\bibinfo {volume} {370}},\ \bibinfo {pages} {238}
  (\bibinfo {year} {2010})}\BibitemShut {NoStop}%
\bibitem [{\citenamefont {H{\"a}nggi}, \citenamefont {Talkner},\ and\
  \citenamefont {Borkovec}(1990)}]{50yKramers}%
  \BibitemOpen
  \bibfield  {author} {\bibinfo {author} {\bibfnamefont {P.}~\bibnamefont
  {H{\"a}nggi}}, \bibinfo {author} {\bibfnamefont {P.}~\bibnamefont {Talkner}},
  \ and\ \bibinfo {author} {\bibfnamefont {M.}~\bibnamefont {Borkovec}},\
  }\href@noop {} {\bibfield  {journal} {\bibinfo  {journal} {Rev. Mod. Phys.}\
  }\textbf {\bibinfo {volume} {62}},\ \bibinfo {pages} {2} (\bibinfo {year}
  {1990})}, {also see references within} \BibitemShut {NoStop}%
\bibitem [{\citenamefont {Fuchs}(1964)}]{Fuchs}%
  \BibitemOpen
  \bibfield  {author} {\bibinfo {author} {\bibfnamefont {N.}~\bibnamefont
  {Fuchs}},\ }\href@noop {} {\emph {\bibinfo {title} {The Mechanics of
  Aerosols}}}\ (\bibinfo  {publisher} {Pergamon Press,Oxford},\ \bibinfo {year}
  {1964})\BibitemShut {NoStop}%
\bibitem [{\citenamefont {Berezhkovskii}\ and\ \citenamefont
  {Szabo}(2011)}]{Berezhkovskii2011}%
  \BibitemOpen
  \bibfield  {author} {\bibinfo {author} {\bibfnamefont {A.~M.}\ \bibnamefont
  {Berezhkovskii}}\ and\ \bibinfo {author} {\bibfnamefont {A.}~\bibnamefont
  {Szabo}},\ }\href {\doibase 10.1063/1.3626215} {\bibfield  {journal}
  {\bibinfo  {journal} {J. Chem. Phys.}\ }\textbf {\bibinfo {volume} {135}},\
  \bibinfo {pages} {074108} (\bibinfo {year} {2011})}\BibitemShut {NoStop}%
\bibitem [{\citenamefont {Burada}\ \emph {et~al.}(2007)\citenamefont {Burada},
  \citenamefont {Schmid}, \citenamefont {Reguera}, \citenamefont {Rub\'\i},\ and\
  \citenamefont {H{\"a}nggi}}]{Burada2007}%
  \BibitemOpen
  \bibfield  {author} {\bibinfo {author} {\bibfnamefont {P.~S.}\ \bibnamefont
  {Burada}}, \bibinfo {author} {\bibfnamefont {G.}~\bibnamefont {Schmid}},
  \bibinfo {author} {\bibfnamefont {D.}~\bibnamefont {Reguera}}, \bibinfo
  {author} {\bibfnamefont {J.~M.}\ \bibnamefont {Rub\'\i}}, \ and\ \bibinfo
  {author} {\bibfnamefont {P.}~\bibnamefont {H{\"a}nggi}},\ }\href@noop {}
  {\bibfield  {journal} {\bibinfo  {journal} {Phys. Rev. E}\ }\textbf {\bibinfo
  {volume} {75}},\ \bibinfo {pages} {051111} (\bibinfo {year}
  {2007})}\BibitemShut {NoStop}%
\bibitem [{\citenamefont {Kalinay}\ and\ \citenamefont
  {Percus}(2006)}]{Kalinay2006}%
  \BibitemOpen
  \bibfield  {author} {\bibinfo {author} {\bibfnamefont {P.}~\bibnamefont
  {Kalinay}}\ and\ \bibinfo {author} {\bibfnamefont {J.~K.}\ \bibnamefont
  {Percus}},\ }\href@noop {} {\bibfield  {journal} {\bibinfo  {journal} {Phys.
  Rev. E}\ }\textbf {\bibinfo {volume} {74}},\ \bibinfo {pages} {041203}
  (\bibinfo {year} {2006})}\BibitemShut {NoStop}%
\bibitem [{\citenamefont {Sokolov}(2010)}]{Sokolov2010}%
  \BibitemOpen
  \bibfield  {author} {\bibinfo {author} {\bibfnamefont {I.~M.}\ \bibnamefont
  {Sokolov}},\ }\href@noop {} {\bibfield  {journal} {\bibinfo  {journal} {Eur.
  J. Phys.}\ }\textbf {\bibinfo {volume} {31}},\ \bibinfo {pages} {1353}
  (\bibinfo {year} {2010})}\BibitemShut {NoStop}%
\bibitem [{\citenamefont {Harayama}, \citenamefont {Klages},\ and\
  \citenamefont {Gaspard}(2002)}]{Gaspard2002}%
  \BibitemOpen
  \bibfield  {author} {\bibinfo {author} {\bibfnamefont {T.}~\bibnamefont
  {Harayama}}, \bibinfo {author} {\bibfnamefont {R.}~\bibnamefont {Klages}}, \
  and\ \bibinfo {author} {\bibfnamefont {P.}~\bibnamefont {Gaspard}},\ }\href
  {\doibase 10.1103/PhysRevE.66.026211} {\bibfield  {journal} {\bibinfo
  {journal} {Phys. Rev. E}\ }\textbf {\bibinfo {volume} {66}},\ \bibinfo
  {pages} {026211} (\bibinfo {year} {2002})}\BibitemShut {NoStop}%
\bibitem [{\citenamefont {Cecconi}\ \emph {et~al.}(2003)\citenamefont
  {Cecconi}, \citenamefont {del Castillo-Negrete}, \citenamefont {Falcioni},\
  and\ \citenamefont {Vulpiani}}]{Cecconi2003}%
  \BibitemOpen
  \bibfield  {author} {\bibinfo {author} {\bibfnamefont {F.}~\bibnamefont
  {Cecconi}}, \bibinfo {author} {\bibfnamefont {D.}~\bibnamefont {del
  Castillo-Negrete}}, \bibinfo {author} {\bibfnamefont {M.}~\bibnamefont
  {Falcioni}}, \ and\ \bibinfo {author} {\bibfnamefont {A.}~\bibnamefont
  {Vulpiani}},\ }\href {\doibase 10.1016/S0167-2789(03)00051-4} {\bibfield
  {journal} {\bibinfo  {journal} {Physica D}\ }\textbf {\bibinfo {volume}
  {180}},\ \bibinfo {pages} {129 } (\bibinfo {year} {2003})}\BibitemShut
  {NoStop}%
\end{thebibliography}
\end{document}